\documentclass[twocolumn,secnumarabic,amssymb,nobibnotes,aps,pra]{revtex4-1}

\usepackage{bm}
\usepackage{graphicx}
\usepackage{color}
\usepackage{tabularx}
\usepackage{amsmath}

\usepackage{hyperref}
\hypersetup{breaklinks=true}
\begin{document}

\title{Relativistic many-body analysis of the electric dipole moment enhancement factor of $^{210}$Fr and associated properties}

\author{Nanako Shitara$^1$, B. K. Sahoo$^2$, T. Watanabe$^3$, and B. P. Das$^1$}
\email[Email: ]{das.b.aa@m.titech.ac.jp}
\affiliation{$^1$Department of Physics, School of Science, Tokyo Institute of Technology, Ookayama, Meguro-ku, Tokyo 152-8550, Japan \\
$^2$Atomic, Molecular and Optical Physics Division, Physical Research Laboratory, Navrangpura, Ahmedabad 380009, India \\
$^3$Global Scientific Information and Computing Center, Tokyo Institute of Technology, Ookayama, Meguro-ku, Tokyo 152-8550, Japan}
\date{\today}

\begin{abstract}
The relativistic coupled-cluster (RCC) method is a powerful many-body method, particularly in the evaluation of electronic wave functions of heavy atoms and molecules, and can be used to calculate various atomic and molecular properties. One such atomic property is the enhancement factor ($R$) of the atomic electric dipole moment (EDM) due to an electron EDM  needed in electron EDM searches. The EDM of the electron is a sensitive probe of CP-violation, and its search could provide insights into new physics beyond the Standard Model, as well as open questions in cosmology. Electron EDM searches using atoms require the theoretical evaluation of $R$ to provide an upper limit for the magnitude of the electron EDM. In this work, we calculate $R$ of $^{210}$Fr in the ground state using an improved RCC method, and perform an analysis on the many-body processes occurring within the system. 
The RCC method allows one to capture the effects of both the electromagnetic interaction and P- and T-violating interactions, and our work develops this method beyond what had been implemented in the previous works. We also perform calculations of hyperfine structure constants, electric dipole transition matrix elements, and excitation energies, to assess the accuracy of $R$ and the success of our improved method. Finally, we present calculations of $R$ with corrections due to Breit interaction effects, approximate quantum electrodynamics (QED) effects, and some leading triple excitation terms added perturbatively, to assess how significantly these terms contribute to the result. We obtain a final value of $R = 799$, with an estimated 3\% error, which is about 11\% smaller than a previously reported theoretical calculation.
\end{abstract}

\maketitle

\section{Introduction}
A consideration of both relativistic and many-body theories is necessary to sufficiently describe heavy atomic and molecular systems~\cite{DasRev, HandbookRQC}. Although there are many theoretical approaches to the evaluation of heavy atomic and molecular wave functions, the relativistic coupled-cluster (RCC) method is to date the method of choice for high-accuracy calculations~\cite{Sato}. The RCC method takes the Dirac-Fock (DF) wave function as its starting point, then considers particle-hole excitations from it, to take electron correlation effects into consideration, and is able to simultaneously account for relativistic and many-body effects. It is a powerful many-body method which has been applied in a variety of fields, including nuclear physics~\cite{Dean, Hagen} and condensed-matter physics~\cite{Bishop}. The RCC method adapted to lattices~\cite{Bishop} could also be applied to optical physics and quantum information. Its advantage over other post-DF methods comes from its ability to include correlation effects to all orders perturbatively for all levels of particle-hole excitation. It also has the property that it is size-extensive~\cite{Bishop}. However, the computational cost scales rapidly with system size, especially compared with other, more approximate post-DF methods. Nevertheless, the power that the coupled-cluster method in general provides in determining many-body wave functions to high accuracy has earned it the reputation as the ``gold standard of many-body methods''~\cite{Thom}. There have been many works~\cite{Sahoo, Sahoo2, Sahoo3, Sakurai, Prasannaa, Abe, Cheng, Shee} which have applied various implementations of the coupled-cluster method to the calculations of different atomic and molecular properties, but all have had to introduce approximations to reduce the unfeasible computational cost associated with a full calculation~\cite{HandbookRQC}. In this work, we apply an improved RCC method in the evaluation of a physical quantity of significance in fundamental physics: the enhancement factor of the electric dipole moment of atomic francium, towards the search for an electron electric dipole moment. In addition, calculations of selected measured properties, such as the magnetic dipole hyperfine constants and the electric dipole transition amplitudes, have been performed. There have been high-precision spectroscopic measurements made on various properties of francium~\cite{Grossman, Simsarian, Sansonetti}, and the application of our new RCC method to the calculation of these measured quantities will allow us to test the strength of the method in giving reliable results.

For all its successes in describing fundamental particle interactions, the Standard Model (SM) of particle physics is unable to account for some outstanding observations in the universe. One of these is the so-called baryon asymmetry in the universe, where the matter-antimatter ratio observed in the current universe is off by several orders of magnitude, compared to the value predicted using SM~\cite{Canetti}. One of the proposed reasons for this discrepancy is the need for additional sources of CP-violation, which is the combined violation of charge conjugation (C) and parity (P) symmetries~\cite{Canetti}. A signature of CP-violation not yet observed is the intrinsic electric dipole moment (EDM) of the electron. For a particle to possess an intrinsic EDM, both P and time-reversal (T) symmetries must be independently violated~\cite{Landau, Ballentine, Sandars1, Kriplovich}. From the CPT theorem, T-violation implies CP-violation. 
The SM predicts an electron EDM value, denoted by $d_e$, of $\vert d_e \vert \sim 10^{-38}$ e-cm~\cite{Bernreuther, Chupp}, which falls far outside the range of values current experiments can probe. However, several beyond the SM (BSM) paradigms predict values of the electron EDM that are many orders of magnitude larger, such as variants of the Supersymmetric (SUSY) model and the left-right symmetric model, which, depending on the parameters, can predict a value of the electron EDM as large as $\vert d_e \vert \sim10^{-27}$ e-cm~\cite{Bernreuther, Cirigliano}, which is within reach of current experiments. Thus, a successful measurement of a nonzero electron EDM would provide direct evidence for BSM physics~\cite{Cesarotti}, as well as shed insight into the observed baryon asymmetry in the universe~\cite{Fuyuto}.
Even if the electron EDM is not observed, imposing upper limits on its magnitude can constrain BSM models, which predict different ranges of possible electron EDM values~\cite{Fukuyama}.

The observation of the electron EDM has eluded experimentalists for over half a century, but not without significant improvements to the upper limits. Currently, heavy open-shell atoms and polar molecules are the most promising systems with which to determine the upper bounds on the magnitude of the electron EDM, with the best limit to date set by experiments on thorium oxide (ThO), at $\vert d_e \vert < 1.1\times10^{-29}$ e-cm with 90\% confidence~\cite{ACME}. The best experimental limit using atoms comes from $^{205}$Tl, at $\vert d_e \vert \leq 1.6 \times 10^{-27}$ e-cm with 90\% confidence~\cite{Regan}. Because the electron EDM is known to be extremely small, high precision is required for these experiments. In the case of atomic systems, the presence of a permanent electron EDM can induce an atomic EDM. This atomic EDM can be many times larger than the magnitude of the electron EDM for some systems~\cite{Sandars2}. It is this enhanced EDM that is exploited in electron EDM experiments using atoms. The energy shift due to the atomic EDM is measured in experiments. To obtain an upper limit for the magnitude of the electron EDM from this quantity, the enhancement factor $R$, defined as the ratio of the atomic EDM to the electron EDM, must be theoretically evaluated. 

In this work, we calculate $R$ of the atomic EDM of $^{210}$Fr in the ground state. In many respects, Fr is a suitable candidate for an EDM experiment. Fr is the heaviest alkali atom, which means that it is a highly relativistic system and that it has a single valence electron, making it the atom with the highest predicted EDM enhancement factor out of all atomic candidates on which electron EDM search experiments are currently being performed. Its projected sensitivity is about two orders of magnitude better than the limit given by $^{205}$Tl~\cite{HaradaFPUA}. The greatest advantage of Fr over the other electron EDM search candidates explored in the past is that many isotopes can be prepared~\cite{Kawamura1}, on which EDM experiments can be done separately. This allows for the detection of signatures of CP-violating sources other than the electron EDM, in particular, the scalar-pseudoscalar (S-PS) interaction. In order to comprehensively study BSM physics, the S-PS interaction term must also be considered when performing EDM experiments. 
So far, there have not been as many comprehensive studies on the contribution of the S-PS interaction to atomic EDM, even though it must be considered if the electron EDM is measured. As a more general point, the advantage of atomic systems over molecular candidates is the ability for theoretical calculations to be evaluated with a higher accuracy due to its simpler electronic structure, and thus the ability to obtain limits to higher accuracy. Furthermore, using different isotopes of the same atom to evaluate the coupling constants for the S-PS interaction allows systematic errors from experiments to be reduced, compared to performing the same measurements on different molecular species. For these reasons, a re-investigation into EDM enhancement of Fr will be of value to the ongoing search for the electron EDM. Electron EDM search experiments using Fr are currently in progress at the University of Tokyo, further motivating this study~\cite{CNS,Sakemi}. Another electron EDM search experiment using $^{211}$Fr has also been proposed by Wundt et al.~\cite{Wundt} and Munger et al.~\cite{Munger}, at TRIUMF in Canada. In this work we focus on the properties of $^{210}$Fr, because the experiments at the University of Tokyo will be using this isotope for the electron EDM search experiments.

The enhancement factor is calculated by numerically evaluating the wave function of the many-body electronic state of the atom using an improved RCC method. The wave function is then used to evaluate the appropriate expectation value. The contribution of individual RCC terms are analyzed and discussed, particularly in relation to the specific many-body effects they contain. The application of the RCC method to the calculation of atomic EDM was first proposed in 1994 by Shukla et al.~\cite{Shukla}, and was implemented for open-shell atoms for the first time in 2008 by Nataraj et al.~\cite{Nataraj}. Calculations on Fr have been performed in the past, first by Sandars in 1966~\cite{Fr3}, using a one electron central force potential approximation, and later by Byrnes et al. in 1999~\cite{Fr2}, using a sum-over-states approach, and by Mukherjee et al. in 2009~\cite{Fr1}, using an approximate RCC method. Our work aims to advance these past results, by using an improved RCC method that addresses the weaknesses of the previous RCC calculation by Mukherjee et al.~\cite{Fr1}. The implemented upgrades include an improved basis set and the inclusion of terms that were omitted in previous calculations~\cite{Fr1,Fr2} due to computational cost limitations. In view of the recent progress in ongoing Fr EDM experiments~\cite{HaradaFPUA}, evaluating an improved theoretical result is of importance in yielding a limit for the electron EDM value and related quantities. High-performance computing is utilized in these calculations to include as many terms as possible for improving accuracy, as well as to include correction terms due to physical effects, such as the Breit interaction~\cite{Grant} and quantum electrodynamic (QED) effects~\cite{Shabaev}, which have not been considered in previous Fr EDM calculations. A subset of triple excitation terms were also evaluated perturbatively and its contribution added to the result. Finally, the accuracies of the results are assessed by comparing various physical quantities evaluated using the calculated wave function with their corresponding experimental values, to ensure that the quality of the RCC state is sufficiently good. In particular, we have calculated the hyperfine structure constants, electric dipole (E1) transition matrix elements, and excitation energies of selected states of $^{210}$Fr. For the hyperfine structure constants and the E1 transition matrix elements, we have included various correction terms to enhance the accuracies of the results and compared these against available experimental and other theoretical values. These results serve to highlight the strengths of our improved RCC method, implemented on open-shell atoms for the first time in this work.

\section{Theory}
We start with the Dirac-Coulomb (DC) Hamiltonian~\cite{Dirac} of an atomic system, given in atomic units as
\begin{equation}
\hat{H}_0 = \sum_i \big(c \bm{\alpha} \cdot \bm{p}_i + (\beta-1) c^2 + V_\mathrm{nuc}(r_i)\big) + \sum_{i<j} \frac{1}{r_{ij}},
\end{equation}
where summations are taken over electrons $i$ and pairs of electrons $i,j$ in the atom, respectively, $c$ is the speed of light, $\bm{\alpha}$ and $\beta$ are Dirac matrices, $\bm{p}_i$ is the momentum operator for electron $i$, and $V_\mathrm{nuc}$ is the potential due to the atomic nucleus. $\frac{1}{r_{ij}}$ is the Coulomb operator, where $r_{ij}$ refers to the distance between electrons $i$ and $j$.
If we assume that a nonzero electron EDM exists, a term corresponding to the interaction of the electron EDM with the internal electric field of the atom must be added to the DC Hamiltonian, and the resulting atomic Hamiltonian $\hat{H}$ becomes
\begin{equation}
\hat{H} = \hat{H}_0 + d_e\hat{H}',
\end{equation}
where $H'$ is a P- and T-violating perturbation to the Hamiltonian, and has the expression
\begin{equation}
\hat{H}' = -\beta\bm{\Sigma}\cdot \bm{\mathcal{E}}^\mathrm{int}, \label{eq:Hpert}
\end{equation}
with $\bm{\mathcal{E}}^\mathrm{int}$ denoting the internal electric field of the atom and $\bm{\Sigma}$ defined by 
\begin{equation}
{\bm \Sigma} = \begin{pmatrix}
{\bm \sigma} & 0 \\
0 & {\bm \sigma}
\end{pmatrix}\label{eq:SigmaDef}
\end{equation} 
where $\bm{\sigma}$ are the Pauli spin matrices.
$d_e$ is small, so the electron EDM interaction term can be treated as a perturbation to the DC Hamiltonian, with $d_e$ taken as the perturbation parameter.
The atomic wave function $\vert \Psi_\alpha \rangle$ is then expressed as a first order perturbed wave function whose unperturbed component satisfies the many-electron Dirac equation using the DC Hamiltonian $\hat{H}_0$, and the first order perturbation term is evaluated through standard perturbation theory by treating the electron EDM interaction operator as the perturbation {to the DC Hamiltonian}. That is,
\begin{equation}
\vert \Psi_\alpha \rangle \approx \vert \Psi_\alpha^{(0)} \rangle + d_e \vert \Psi_\alpha^{(1)} \rangle, \label{eq:perturbedWF}
\end{equation}
where the equation
\begin{equation}
\hat{H}_0\vert\Psi_\alpha^\mathrm{(0)}\rangle = E_0^{(0)}\vert \Psi_\alpha^\mathrm{(0)}\rangle
\end{equation}
is satisfied.
Note that, in Eq.~(\ref{eq:perturbedWF}), the first order perturbed wave function is $d_e\vert \Psi_\alpha^{(1)} \rangle$, not $\vert \Psi_\alpha^{(1)} \rangle$.

Our aim is to obtain an expression for the enhancement factor $R$ of the atomic EDM due to the existence of a nonzero electron EDM. $R$ is defined like
\begin{equation}
R = \frac{\langle D_a \rangle}{d_e}, \label{eq:R}
\end{equation}
where $\langle D_a \rangle$ is the magnitude of the atomic EDM, and $d_e$ is the magnitude of the electron EDM. By definition, an EDM induces an energy shift $\Delta E$ that is linear in the magnitude of the applied electric field $\mathcal{E}$:
\begin{equation}
\Delta E = -\langle D_a \rangle \mathcal{E}. \label{eq:shiftE}
\end{equation}
We derive an expression for the atomic EDM induced by an electron EDM and an external electric field. This is given as the normalized expectation value of the atomic dipole operator $\bm{D}_a$ with respect to the atomic state of interest $\vert \Psi \rangle$, like
\begin{equation}
\langle \bm{D}_a \rangle  = \frac{\langle \Psi \vert \bm{D}_a \vert \Psi \rangle}{\langle \Psi \vert \Psi \rangle}. \label{eq:aEDMexpr}
\end{equation}
In this case, we are interested in the atomic ground state which we denote as $\vert \Psi_\alpha \rangle$.
In the presence of an external electric field and a nonzero electron EDM, the dipole operator takes the form
\begin{eqnarray}
\bm{D}_a &=& e\bm{r} + d_e \beta \bm{\Sigma} \\
&=& \bm{D} + d_e \beta \bm{\Sigma},
\end{eqnarray}
where the first term is due to the EDM induced by the external field, and the second term due to the electron EDM. 
$\beta$ is the Dirac matrix, and $\bm{\Sigma}$ is defined in Eq.~(\ref{eq:SigmaDef}). The summation over each electron is suppressed.

Substituting Eq.~(\ref{eq:perturbedWF}) into the numerator of Eq.~(\ref{eq:aEDMexpr}),
\begin{eqnarray}
\langle \Psi_\alpha \vert \bm{D}_a \vert \Psi_\alpha \rangle &=& (\langle \Psi_\alpha^{(0)} \vert + d_e \langle \Psi_\alpha^{(1)} \vert)\bm{D}_a(\vert \Psi_\alpha^{(0)} \rangle + d_e \vert \Psi_\alpha^{(1)} \rangle) \nonumber \\
&=& \langle \Psi_\alpha^{(0)} \vert \bm{D}_a \vert \Psi_\alpha^{(0)} \rangle +2 d_e \langle \Psi_\alpha^{(0)} \vert \bm{D}_a \vert \Psi_\alpha^{(1)} \rangle \label{eq:aEDMexpand}
\end{eqnarray}
keeping only terms linear in $d_e$.
Now, we note that atomic wave functions have well-defined parities. The DC Hamiltonian is parity conserving, but the perturbation introduced, which is the electron EDM interaction term, is parity violating. Thus, $\vert \Psi_\alpha^{(1)} \rangle$ and $\vert \Psi_\alpha^{(0)} \rangle$ have opposite parities. Noting the fact that $\bm{D}$ is an odd parity operator and $d_e \beta \bm{\Sigma}$ is an even parity operator, Eq.~(\ref{eq:aEDMexpand}) can be simplified into
\begin{equation}
\langle \Psi_\alpha \vert \bm{D}_a \vert \Psi_\alpha \rangle = \langle \Psi_\alpha^\mathrm{(0)} \vert d_e \beta \bm{\Sigma} \vert \Psi_\alpha^\mathrm{(0)} \rangle + 
2 d_e \langle \Psi_\alpha^\mathrm{(1)} \vert \bm{D}  \vert \Psi_\alpha^\mathrm{(0)} \rangle \label{eq:aEDMsimplified}
\end{equation}
using parity selection rules. This is the non-normalized expression for the atomic EDM. Dividing this through by $d_e$ gives us the expression for $R$:
\begin{equation}
R = \frac {\langle \Psi_\alpha^\mathrm{(0)} \vert \beta \bm{\Sigma} \vert \Psi_\alpha^\mathrm{(0)} \rangle + 
2 \langle \Psi_\alpha^\mathrm{(1)} \vert \bm{D}  \vert \Psi_\alpha^\mathrm{(0)} \rangle}{\langle \Psi_\alpha \vert \Psi_\alpha \rangle},
\end{equation}
where $\vert \Psi_\alpha^{(1)} \rangle$ can be expressed like
\begin{equation}
\vert \Psi_\alpha^\mathrm{(1)} \rangle = \sum_{\nu \neq \alpha} \frac{\vert\Psi_\nu^\mathrm{(0)}\rangle \langle\Psi_\nu^\mathrm{(0)}\vert -\beta\bm{\Sigma}\cdot \bm{\mathcal{E}}^\mathrm{int} \vert\Psi_\alpha^\mathrm{(0)}\rangle }{E_\alpha^\mathrm{(0)} - E_\nu^\mathrm{(0)}}
\end{equation}
from perturbation theory, where $k$ denotes all intermediate states.
This expression can be simplified using Dirac algebra like~\cite{Das1}
\begin{eqnarray}
R &=& \frac{2 i c}{\langle \Psi_\alpha \vert \Psi_\alpha \rangle} \times \notag \\
&& \sum_{\nu \neq \alpha} \left( \! \frac{\langle \Psi_\alpha^\mathrm{(0)}\vert \beta\gamma_5{p}^2 \vert \Psi_\nu^\mathrm{(0)}\rangle \langle \Psi_\nu^\mathrm{(0)}\vert  D  \vert \Psi_\alpha^\mathrm{(0)}\rangle}{{E}_\alpha^\mathrm{(0)}  - {E}_\nu^\mathrm{(0)}} \! + \! \mathrm{h.c.} \! \right) \! \label{eq:R-expr} \\
&=& \frac{2 \langle \Psi_\alpha^{\mathrm{(1)}'} \vert D \vert \Psi_\alpha^\mathrm{(0)} \rangle}{\langle \Psi_\alpha \vert \Psi_\alpha \rangle} \label{eq:R-expr-short}
\end{eqnarray}
where h.c. denotes the Hermitian conjugate of the preceding term, and 
\begin{equation}
\vert \Psi_\alpha^{\mathrm{(1)}'} \rangle = \sum_{\nu \neq \alpha} \frac{\vert\Psi_\nu^\mathrm{(0)}\rangle \langle\Psi_\nu^\mathrm{(0)}\vert 2ic\beta\gamma_5 p^2 \vert\Psi_\alpha^\mathrm{(0)}\rangle }{E_\alpha^\mathrm{(0)} - E_\nu^\mathrm{(0)}}.
\end{equation}

We see from Eq.~(\ref{eq:R}) and Eq.~(\ref{eq:shiftE}) that
\begin{equation}
\Delta E =  -d_e R \mathcal{E},
\end{equation}
which shows that we can obtain an upper limit for $d_e$ through the combination of experimental $\Delta E$ value and theoretical calculation of $R$.

\section{Method of calculation}

The RCC method takes as its starting point the Dirac-Fock (DF) state $\vert \Phi_v \rangle$, constructed as a Slater determinant of single-electron wave functions~\cite{Grant}. 
The relativistic single-electron wave functions $\vert \phi \rangle$ have the form 
\begin{equation}
\vert \phi \rangle = \begin{pmatrix} P(r)\chi_{j m_j l_L} \\ i Q(r) \chi_{j m_j l_S} \end{pmatrix}
\end{equation}
where the upper and lower components indicate the large and small components of the relativistic wave function, respectively, $P(r)$ and $Q(r)$ denote the radial parts of each component, and the $\chi$'s denote the spin angular parts of each component which depend on the quantum numbers $j$, $m_j$, and $l$. $l_L$ denotes $l$ for the large component, while $l_S$ denotes $l$ for the small component. The spin quantum number $s$ is fixed at $s=1/2$, because we are only considering electrons here.

The radial parts of the large and small components of the relativistic single-electron wave functions are constructed as a sum of Gaussian functions, called Gaussian type orbitals (GTO's). For a given $l$ and $j$, the large and small components of the radial wave functions have the form
\begin{eqnarray}
P(r) &=& \sum^N_{i=1} c_i^L g_i^L(r) \\
\text{and} \quad Q(r) &=& \sum^N_{i=1} c_i^S g_i^S(r),
\end{eqnarray}
respectively, where $c_i$ is the coefficient for orbital $i$, and the superscript $L$($S$) refers to the large (small) component of the relativistic wave function. The small component is evaluated from the large component using the kinetic balance condition~\cite{Dyall1}, like
\begin{equation}
(\bm{\sigma}\cdot\bm{p})g_i^L = g_i^S.
\end{equation}
The GTO of the large component for a given $l$ and $j$ is given as
\begin{equation}
g_i^{L}(r) = r^{l} e^{-\alpha_i r^2},
\end{equation}
where the exponent $\alpha_i$ is given by
\begin{equation}
\alpha_i = \alpha_0 \beta^{i-1}.
\end{equation}
This condition on the exponent is called the even-tempered condition~\cite{Quiney}. The parameters $\alpha_0$ and $\beta$ are optimized for each angular symmetry, and the values used in this calculation are as shown in Table~\ref{tb:params}. The optimization was performed so that the bound orbital energies and the expectation values of $r$, $1/r$, and $1/r^2$ matched those obtained through a direct differential equation method using the \texttt{GRASP} code~\cite{Dyall2}. The differential equation DF calculation employed in \texttt{GRASP} does not rely on any external parameters, but is unable to  provide continuum orbitals, while the matrix DF calculation that we employ is able to do this. Therefore, by ensuring that the bound orbitals we obtain for some choice of parameters give similar expectation values to those obtained using \texttt{GRASP}, we ensure that our choice of parameters is reasonable. The value of $R$ in the DF atomic ground state and the values of the magnetic dipole hyperfine interaction constants in selected DF atomic states were also ensured to match the values evaluated using DF states constructed from \texttt{GRASP} orbitals. The range of orbitals calculated at the DF level, and the range of active orbitals used in the RCC calculation, are also shown in the same table, for each symmetry. \\

\begin{table}[h!]
\caption{The optimized values of $\alpha_0$ and $\beta$, the range of the principal quantum numbers of the orbitals calculated using the DF method ($n_\mathrm{DF}$), and the range of the principal quantum numbers of the DF orbitals used in the RCC calculation ($n_\mathrm{RCC}$) for each angular symmetry used in this work.}

 \begin{tabularx}{8.6cm}{X | X X X X X X} 
 \hline
 \hline
  & $s$ & $p$ & $d$ & $f$ & $g$ \\ [0.5ex] 
 \hline
 $\alpha_0$ & 0.0009  & 0.0008 & 0.001 & 0.004 & 0.005 \\
 $\beta$ & 2.25  & 2.20  & 2.15  & 2.25 & 2.35 \\
 $n_\mathrm{DF}$ & 1-40 & 2-40 & 3-40 & 4-40 & 5-40 \\
 $n_\mathrm{RCC}$ & 1-20 & 2-21 & 3-22 & 4-20 & 5-20 \\
 \hline 
 \hline
\end{tabularx}

\label{tb:params}
\end{table}

The RCC wave function $\vert \Psi_\alpha^{(0)} \rangle$ of a particular atomic state is constructed as a linear combination of $n$ particle-$n$ hole excitations of the DF state and the DF state, and expressed like
\begin{equation}
\vert \Psi_\alpha^{(0)} \rangle = e^{T^{(0)}}(1+S^{(0)})\vert \Phi_v \rangle, \label{eq:RCCwfn}
\end{equation}
where $T^{(0)}$ is the sum of all possible $n$ particle-$n$ core orbital excitation operators ($T^{(0)} = \sum_n T_n^{(0)}$), $S^{(0)}$ is the sum of all possible $n$ particle-$n$ valence orbital excitation operators ($S^{(0)} = \sum_n S_n^{(0)}$), and $\vert \Phi_v \rangle$ is the DF state, where the valence orbital is treated like a particle orbital; that is, $\vert \Phi_v \rangle = a_v^\dagger \vert \Phi_0 \rangle$, where $\vert \Phi_0 \rangle$ consists of occupied core orbitals. This expression for $\vert \Psi_\alpha^{(0)} \rangle$ means that in general, it is not normalized. It should be noted that in the DF calculation, the orbital wave functions are evaluated assuming a $V_{N-1}$ potential, outlined in Ref.~\cite{Kelly} and Ref.~\cite{Das2} and references therein, where $N$ is the atomic number, which is 87 for Fr. It should also be noted that it is these particle-hole excited states that account for the electron-electron correlation effects within the atom, which were neglected in the DF state.

In the presence of the P- and T-violating perturbation to the Hamiltonian, the RCC wave function should take the form given in Eq.~(\ref{eq:perturbedWF}), which should match Eq.~(\ref{eq:RCCwfn}) when the following substitutions are made:
\begin{eqnarray}
T^{(0)} &\rightarrow& T^{(0)} + d_e T^{(1)} \\
\text{and} \quad  S^{(0)} &\rightarrow& S^{(0)} + d_e S^{(1)}.
\end{eqnarray}
Equating terms that are of the same order in $d_e$, we retrieve the expressions for the unperturbed and the first order perturbed RCC wave functions
\begin{eqnarray}
\vert \Psi_\alpha^{(0)} \rangle &=& e^{T^{(0)}}(1+S^{(0)})\vert \Phi_v \rangle \\
\text{and} \quad \vert \Psi_\alpha^{(1)'} \rangle &=& e^{T^{(0)}}(S^{(1)} + T^{(1)} + T^{(1)} S^{(0)})\vert \Phi_v \rangle.
\end{eqnarray}
These satisfy the unperturbed and the first order perturbed many-electron Dirac equations, given respectively as
\begin{eqnarray}
\hat{H}_0\vert\Psi_\alpha^\mathrm{(0)}\rangle &=& E_0\vert \Psi_\alpha^\mathrm{(0)}\rangle \\
\text{and} \quad (\hat{H}_0-E_0)\vert\Psi_\alpha^\mathrm{(1)'}\rangle &=& (E^\mathrm{(1)'} - \hat{H}')\vert \Psi_\alpha^\mathrm{(0)}\rangle.
\end{eqnarray}

The amplitudes for $T^{(0)}$ and $T^{(1)}$ are evaluated by solving the unperturbed and the first order perturbed many-electron Dirac equations that hold for the core electrons, and $S^{(0)}$ and $S^{(1)}$ are evaluated by solving the above two equations in the form given in Ref.~\cite{Sahoo} and references therein. We note here that the perturbed wave function $\vert \Psi_\alpha^{(1)} \rangle$ is evaluated directly by solving the first order perturbed Dirac equation, as opposed to evaluating it as a sum over explicitly constructed intermediate states, as was done by Byrnes et al. in Ref.~\cite{Fr2}. This gives a more accurate expression for the perturbed state, since the approximation introduced by truncating the infinite sum of intermediate states when evaluating this computationally is not necessary in this approach. 

\begin{figure}[h]
\centering
\caption{Goldstone diagrams of the (a) ${D}S_1^{(1)}$, (b) ${D}S_2^{(1)}$, and (c) ${D}T_1^{(1)}$ contributions to the RCCSD expression of the atomic EDM. The h.c. diagrams and the exchange term diagrams are not shown.}
\includegraphics[width=8.6cm]{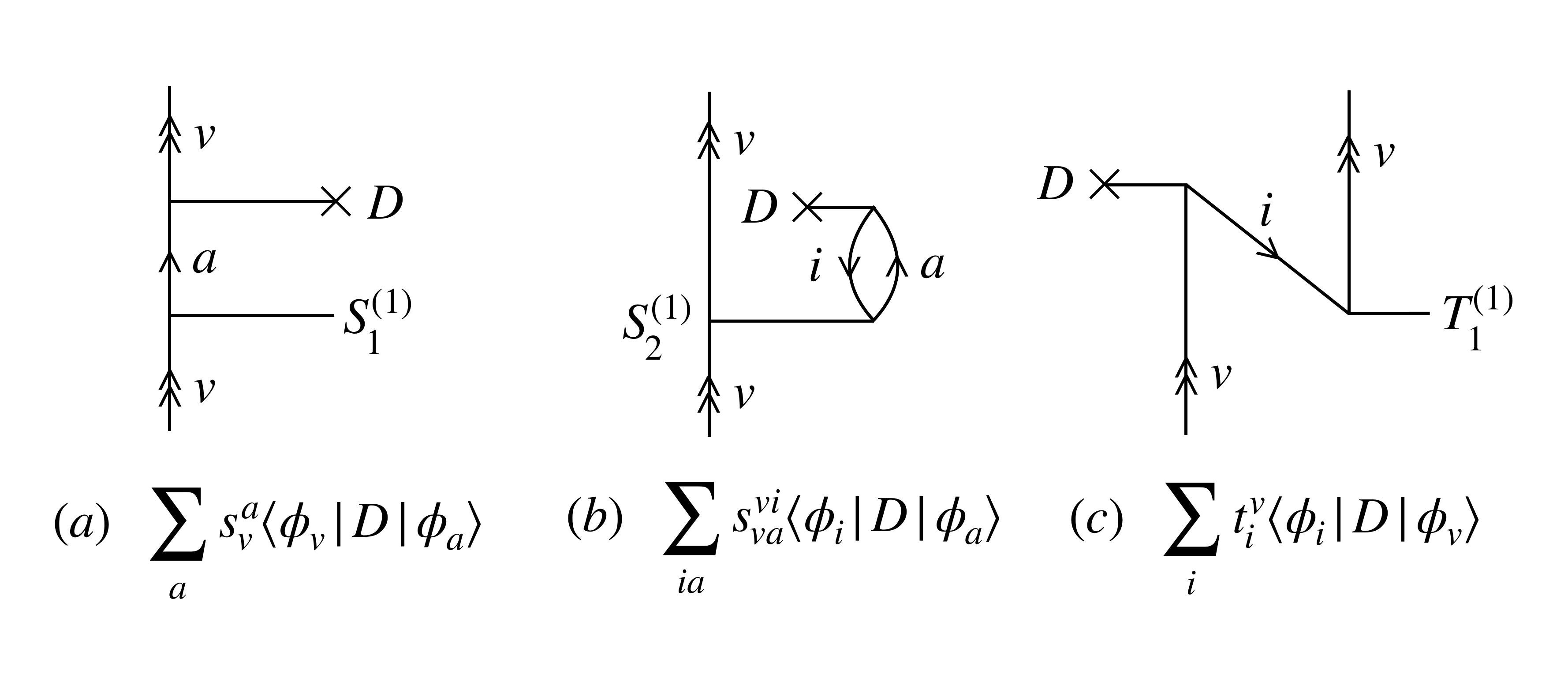}
\label{fig:diagrams}
\end{figure}

\begin{figure}[h]
\centering
\caption{Goldstone diagram representation of the ${D}S_1^{(1)}$ term, expanded in terms of perturbations in the residual interaction. The first term on the right hand side of the equality shows the first order perturbed term in the residual Coulomb interaction, and the second term shows a second order perturbed term in the residual interaction.}
\includegraphics[width=8.6cm]{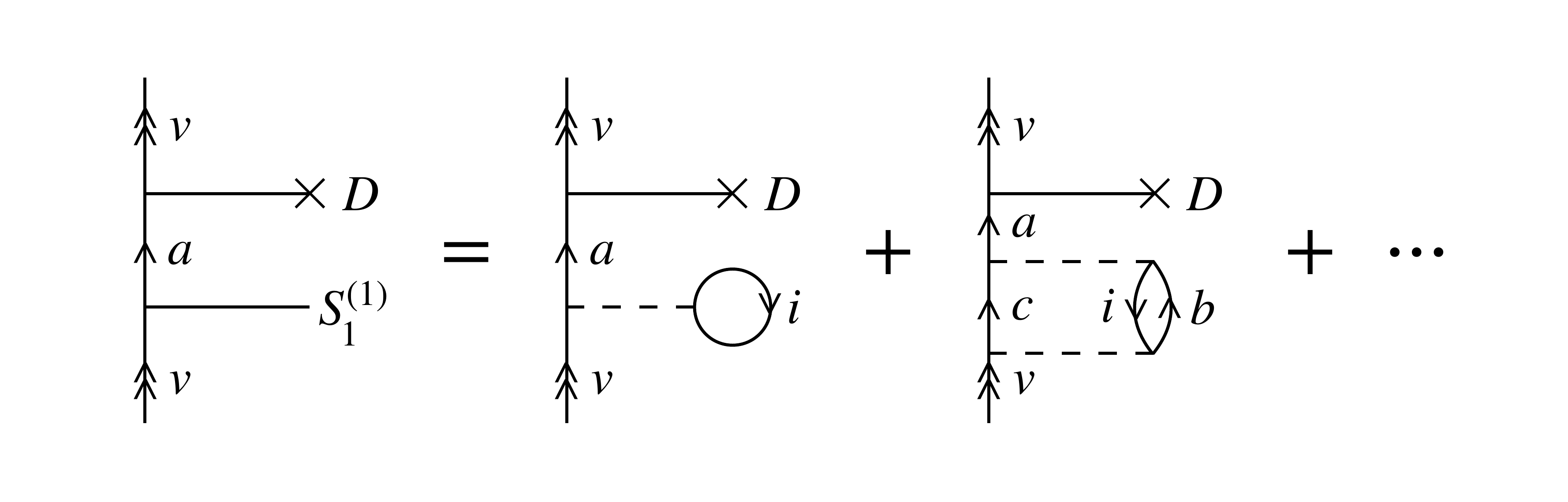}
\label{fig:aExpand}
\end{figure}

Substituting the RCC wave functions in the numerator of Eq.~(\ref{eq:R-expr-short}) and expanding yields
\begin{widetext}
\begin{equation}
\langle \Psi_\alpha^{(1)'} \vert D \vert \Psi_\alpha^{(0)} \rangle = d_e \langle \Phi_v \vert \left( \bar{D} S^{(1)} + \bar{D} T^{(1)} +  \bar{D} T^{(1)} S^{(0)} + {S^{(0)}}^\dagger \bar{D} T^{(1)} + {S^{(0)}}^\dagger \bar{D} S^{(1)} + {S^{(0)}}^\dagger \bar{D} T^{(1)} S^{(0)} \right) + \text{h.c.} \vert \Phi_v \rangle. \label{eq:longD}
\end{equation}
\end{widetext}
where $\bar{D} = e^{{T^{(0)}}^\dagger} D e^{T^{(0)}}$, and taking parity selection rules into account. 
Of the resulting six terms of the numerator and their h.c., the most important terms are $\bar{D}S_1^{(1)}$+ h.c., $\bar{D}S_2^{(1)}$+ h.c., and $\bar{D}T_1^{(1)}$ + h.c. The dominant Goldstone diagrams~\cite{Lindgren} in each of these three terms are shown in Figure~\ref{fig:diagrams}. 
The advantage of this RCC method, sometimes called the expectation value RCC (XRCC) method, is its ability to include perturbations to all orders in the residual interaction, which is the difference between the exact two-body interaction and the DF approximation of the two-body interaction. As an example, we take diagram (a) of Fig.~\ref{fig:diagrams} and expand it in terms of perturbations of the residual interaction in Fig.~\ref{fig:aExpand}. The first diagram on the right hand side of the equality shows the first order perturbed term in the residual Coulomb interaction, and the second diagram a second order perturbed term. Other second order terms and higher order terms are not shown, and are indicated by the ellipsis. Fig.~\ref{fig:aExpand} shows that the RCC term $DS_1^{(1)}$ contains terms of all orders of perturbation in the residual Coulomb interaction. Generally speaking, all RCC terms contain terms of all orders of perturbation in the residual interaction similarly. 
{It can thus be seen that the XRCC approach makes the physical effects transparent through the use of RCC diagrams, which are a compact way of representing larger classes of many-body perturbation theory (MBPT) diagrams.}
These MBPT diagrams show the particular many-body interactions occurring in the excitation process. For example, the Coulomb interactions in the second order perturbed term in Fig.~\ref{fig:aExpand} is known as the Brueckner pair correlation (BPC), which is one type of many-body interaction contained within the post-DF residual interaction~\cite{Brueckner, Nesbet}. Other many-body interaction processes can be similarly identified through the MBPT diagrams.

In this calculation, we employ the RCC singles and doubles (RCCSD) approximation, where we consider only one and two-particle excitations, so that the excitation operators are defined like
\begin{eqnarray}
T^{(0/1)} &=& T^{(0/1)}_1 + T^{(0/1)}_2,\\
\text{and} \quad S^{(0/1)} &=& S^{(0/1)}_1 + S^{(0/1)}_2.
\end{eqnarray}
We have considered excitations of all electrons from the core orbitals in this calculation. 

\section{Results}
The results of our RCC calculation of $R$ of $^{210}$Fr using the DC Hamiltonian, and the DF and leading RCC terms contributing to it, are shown in Table~\ref{tb:DC-210}. Note here that the DF terms are not included in the total sum of the contributions listed, because the DF terms are actually already embedded within the RCC terms listed. Table~\ref{tb:DC-210} also compares our results with the results of a previous RCC calculation of the same atom, and with previous calculations using other many-body methods. We obtain $R = 812$, which is about 10\% less than the value calculated by Mukherjee et al.~\cite{Fr1}. 
\begin{table}[h!]
\caption{DF and the leading RCC contributions to $R$ of the atomic EDM of $^{210}$Fr using the DC Hamiltonian, calculated using the RCCSD method. These values are compared against previous calculations by Mukherjee et al.~\cite{Fr1} (also calculated using the RCCSD method), by Byrnes et al.~\cite{Fr2}, and by Sandars~\cite{Fr3}. ``Norm.'' refers to the correction due to normalization of the RCC wave function, and ``Extra'' refers to contributions due to terms not listed in the table, which have been calculated.}
 \begin{tabularx}{8.6cm}{X  X X  X} 
 \hline
 \hline
 \multicolumn{2}{l}{Terms from RCC theory} & This work & Other~\cite{Fr1}\\ [0.5ex] 
 \hline
 \multicolumn{2}{l}{DF (core)} & 24.85 & 25.77 \\ 
 \multicolumn{2}{l}{DF (valence)} & 702.39 & 695.44 \\
 \multicolumn{2}{l}{$\bar{D}T_1^{(1)} +$ h.c.} & 44.05 & 43.39 \\ 
 \multicolumn{2}{l}{$\bar{D}S_1^{(1)} +$ h.c.} & 889.18 & 1000.19 \\ 
 \multicolumn{2}{l}{$\bar{D}S_2^{(1)} +$ h.c.} & -49.46 & -64.94 \\
 \multicolumn{2}{l}{${S_1^{(0)}}^\dagger \bar{D} S_1^{(1)} +$ h.c.} & -14.15 & -18.07 \\
 \multicolumn{2}{l}{${S_2^{(0)}}^\dagger \bar{D} S_1^{(1)} +$ h.c.} & -48.45 & -59.18 \\
 \multicolumn{2}{l}{${S_1^{(0)}}^\dagger \bar{D} S_2^{(1)} +$ h.c.} & -3.84 & -2.80 \\
 \multicolumn{2}{l}{${S_2^{(0)}}^\dagger \bar{D} S_2^{(1)} +$ h.c.} & 11.99 & 19.26 \\
 Extra & & 4.98 & 1.51 \\
 Norm. & & -22.11 & -24.42 \\
 \hline
 Total & & 812.19 & 894.93 \\
 \hline
 \hline
 \multicolumn{3}{l}{Total from other many-body methods} & \\
 \hline
 Ref.~\cite{Fr2}& & & 910(46) \\
 Ref.~\cite{Fr3}& & & 1150 \\
 \hline
 \hline
\end{tabularx}
\label{tb:DC-210}
\end{table}
We note that while the values of each contribution differ between our result and the previous RCC result, the trend of the relative magnitudes of each contribution remain similar. 
Of the contributing terms listed, the largest term is given by $\bar{D}S_1^{(1)} +$ h.c., whose diagram is indicated in Fig.~\ref{fig:diagrams} (a). This term predominantly contains contributions due to the BPC.
The next largest term is given by $\bar{D}S_2^{(1)} +$ h.c. This term predominantly contains contributions due to electron core polarization (ECP) interactions which are mediated through the Coulomb interaction. This interaction is less important than the BPC, and thus gives a smaller contribution.
Another comparable large term is the ${S_2^{(0)}}^\dagger \bar{D} S_1^{(1)} +$ h.c. term. This has a large contribution because it contains both BPC and ECP contributions.
Finally, $\bar{D}T_1^{(1)} +$ h.c. gives the next leading contribution. This term contains mainly ECP interactions which are mediated through the EDM interaction, unlike the terms in $\bar{D}S_2^{(1)} +$ h.c. $\bar{D}T_1^{(1)} +$ h.c. also contains the core DF contribution. From these results we see the relative importance of BPC and ECP effects on the $R$ of Fr.

The result given by Sandars in Ref.~\cite{Fr3} is evaluated by considering a central force potential experienced by the valence electron due to the nucleus and the core electrons. Thus, his result is very approximate and the differences between his result and later results are not surprising. The discrepancy between our result and that of Byrnes et al.~\cite{Fr2} comes from two differences in the methodology. The first is that Byrnes et al. have used a combination of {\it ab initio} and semi-empirical methods to obtain their result~\cite{Fr2}, while our result is evaluated completely {\it ab initio}. The second is that Byrnes et al. have used an explicit sum over states approach, as mentioned earlier. What is more, they have only calculated singly excited valence intermediate states from $7P_\frac{1}{2}$ to $10P_\frac{1}{2}$ 
and included contributions due to higher lying $P_\frac{1}{2}$ states in an approximate manner~\cite{Fr2}.
In our calculation, we have implicitly considered all intermediate states in the configuration space spanned by our basis, which includes singly excited valence states up to  $21P_{1/2}$, and also core excited states. 
Therefore, their results are closer to our $\bar{D}S_1^{(1)} +$ h.c contribution rather than the total result.

The reason for the discrepancy between our results and those of Mukherjee et al~\cite{Fr1} can be attributed to the differences in the calculation methodology and computational details. There are three main differences to note.
The first is the number of basis orbitals used in the RCC calculation. While we used at most 20 orbitals per symmetry, as shown in Table~\ref{tb:params}, the calculation in Ref.~\cite{Fr1} used only (at most) 14 active orbitals per symmetry. Furthermore, we have employed an even-tempered set of GTO basis functions, whose parameters $\alpha_0$ and $\beta$ differ between different orbital symmetries ($s$, $p$, $d$, $\cdots$) as given in Table~\ref{tb:params}, while Ref.~\cite{Fr1} has used a universal basis set, where the parameters are common to all orbitals. That is, while our current work has specified ten parameters for the basis functions, two for each orbital symmetry, Ref.~\cite{Fr1} has only specified two. This allows us to optimize $\alpha_0$ and $\beta$ for each orbital symmetry, which should behave differently from each other.
The large difference in the $\bar{D}S_1^{(1)} +$ h.c. term, of about 111.01, is mainly due to this difference in the calculation. 
By introducing high-lying virtual states in the $s_\frac{1}{2}$ and the $p_\frac{1}{2}$ symmetries, which have large densities near the nucleus, significant contributions from the EDM matrix element have been added to the single valence excitation term. Note that {this increase in the contribution} is counterbalanced by the energy denominator term in the expression for $R$ given in Eq.~(\ref{eq:R-expr}), so that, at a certain point, the contributions due to high-lying states become negligible. 

The second difference is that our calculation includes amplitudes of all multipoles, as opposed to just amplitudes of multipoles satisfying the even parity channel condition, as done in Ref.~\cite{Fr1}. This is explained below.
For general orbitals $p,q,r,s$, the matrix element of the Coulomb operator $\frac{1}{r_{12}}$, which is a two-body operator, can be expressed like
\begin{eqnarray}
\langle j_p m_p \, j_q m_q &\vert& \frac{1}{r_{12}} \vert j_r m_r \, j_s m_s \rangle \\
= \langle j_p m_p \, j_q m_q &\vert& \sum_{kq} \frac{4\pi}{2k+1} Y_{kq}^*(\theta_1, \, \phi_1) Y_{kq}(\theta_2, \, \phi_2) \nonumber \\
&& \times \frac{r_<^k}{r_>^{k+1}} \vert j_r m_r \, j_s m_s \rangle
\end{eqnarray}
in the $jm$ basis, where $Y_{kq}$ refers to the spherical harmonic functions, $r_<^k$ refers to the smaller of $r_1$ and $r_2$, $r_>^{k+1}$ refers to the larger of $r_1$ and $r_2$,  and $k$ refers to the rank.
From this expression it can be seen that $k$ must satisfy the triangular conditions for $p$ and $r$, and for $q$ and $s$.
For the Coulomb operator, additional constraints 
\begin{eqnarray}
(-1)^{l_p + l_r + k} &=& 1 \label{eq:EPC1}\\
\text{and} \quad (-1)^{l_q + l_s + k} &=& 1 \label{eq:EPC2}
\end{eqnarray}
are derived, where $l$ refers to the orbital angular momentum quantum number. This is due to the fact that the Coulomb interaction is a parity conserving operator. In conjunction with the overall parity selection rule
\begin{equation}
(-1)^{l_p + l_q + l_r + l_s} = 1, \label{eq:paritySelection}
\end{equation}
it can be seen that only a subset of values of $p$, $q$, $r$, $s$, and $k$ satisfy all three equations Eq.~(\ref{eq:EPC1}), Eq.~(\ref{eq:EPC2}), and Eq.~(\ref{eq:paritySelection}), and so for each individual Coulomb interaction, the nonzero contributions come only from this subset.
Now consider the excitation operators $T^{(0)}$ and $S^{(0)}$. The one particle-one hole excitation operators $T^{(0)}_1$ and $S^{(0)}_1$ are rank 0 operators, as seen from the fact that their matrix elements are uniquely identified scalar values~\cite{Geetha}. On the other hand, the two particle-two hole excitation operators $T^{(0)}_2$ and $S^{(0)}_2$ can take on nonzero rank values, since the combination of the angular momenta of two particles allows the state to have multiple $k$ values. For all RCC excitation operators, which each contain Coulomb interactions to all orders, Eq.~(\ref{eq:EPC1}) and Eq.~(\ref{eq:EPC2}) do not apply, and so the values of $p$, $q$, $r$, $s$, and $k$ for which the contribution is nonzero is not restricted to those that are nonzero for the Coulomb interaction. Furthermore, for the perturbed terms of the two-body excitation operators, $T_2^{(1)}$ and $S_2^{(1)}$, Eq.~(\ref{eq:paritySelection}) does not hold as well, because these are parity-violating terms, introduced due to the P- (and T-) violating perturbation to the Hamiltonian. Therefore, in general, all combinations of $p$, $q$, $r$, $s$, and $k$ could give a nonzero contribution to the overall result, including the combinations for which the Coulomb interaction contributions would be zero.
In Ref.~\cite{Fr1}, the even parity channel approximations were employed, which considers only the set of orbitals and $k$ values for which Eq.~(\ref{eq:EPC1}) and Eq.~(\ref{eq:EPC2}) are satisfied, for the $T_2^{(0)}$ and $S_2^{(0)}$ operators, i.e. those for which the contributions to the Coulomb interactions are nonzero. This is on the grounds that these combinations of orbitals and $k$ give the dominant contributions, as discussed in Ref.~\cite{Liu}. However, these approximations were introduced because of limitations in computational resources.  In the absence of computational restrictions, there is no compelling reason for terms that do not satisfy the even parity channel condition to be omitted. Therefore, in this work, we considered contributions due to all combinations of orbitals and $k$.

Finally, our calculation included nonlinear terms in $\bar{D}$, which the calculation in Ref.~\cite{Fr1} had not included. Recall that
\begin{eqnarray}
\bar{D} &=& {e^{T^{(0)}}}^\dagger D e^{T^{(0)}} \\
&=& D + {T^{(0)}}^\dagger D + D T^{(0)} + {T^{(0)}}^\dagger D T^{(0)} + \cdots, \label{eq:Dbarexpand}
\end{eqnarray}
where each $T^{(0)} = T_1^{(0)} + T_2^{(0)}$ and ${T^{(0)}}^\dagger = {T_1^{(0)}}^\dagger + {T_2^{(0)}}^\dagger$. It can be seen that terms in $\bar{D}$ can be grouped into powers of ${T^{(0)}}$ and ${T^{(0)}}^\dagger$. In our work, terms to the $n$th power of ${T^{(0)}}$ and ${T^{(0)}}^\dagger$ were calculated iteratively for increasing $n$, until the difference in the total sum between up to the $n$th and the $(n+1)$th terms was less than a threshold value. This self-consistent method of evaluating $\bar{D}$ ensures that the series given in Eq.~(\ref{eq:Dbarexpand}) converges numerically, and therefore effectively terminates.
In Ref.~\cite{Fr1}, terms nonlinear in ${T^{(0)}}$ and ${T^{(0)}}^\dagger$ in Eq.~(\ref{eq:Dbarexpand}) were not calculated, unlike in our work.
Contributions due to the linear dominant term of three selected RCC terms are given in Table~\ref{tb:lin-DC}. This takes the contributions in $\bar{D}T_1^{(1)}$, $\bar{D}S_1^{(1)}$, and $\bar{D}S_2^{(1)}$ for which $\bar{D} = D$.
\begin{table}[h!]
\caption{RCC calculations of linear contributions to $R$ of $^{210}$Fr of three selected RCC terms, compared against other RCC results given by Ref.~\cite{Fr1}.}
 \begin{tabularx}{8.6cm}{X  X  X} 
 \hline
 \hline
 Terms & This work & Other~\cite{Fr1}\\ [0.5ex] 
 \hline
 ${D}T_1^{(1)}$ & 45.35 & 44.82 \\ 
 ${D}S_1^{(1)}$ & 889.27 & 1000.69 \\
 ${D}S_2^{(1)}$ & -57.29 & -61.28 \\ 
 \hline
 \hline
\end{tabularx}
\label{tb:lin-DC}
\end{table}
It can be seen that the linear term accounts for the majority of the contribution to each term. $DS_2^{(1)}$ shows a slightly larger deviation from $\bar{D}S_2^{(1)}$ compared to the other two terms. This shows that the self-consistent method of evaluation discussed above makes a difference for some leading contributions of $R$. We note that this work is the first to apply this technique on an open-shell atomic system.

To analyze the accuracy of our results, we use our calculated RCC wave functions to evaluate other physical quantities which can be compared against experimental results, and which resemble the terms that comprise the expression for $R$. This analysis gives a quantitative insight into the errors contributing to the result. 
The first quantity to compare is the magnetic dipole hyperfine constant $A$, which can be used to estimate the error in the EDM matrix element in Eq.~(\ref{eq:R-expr}). The error in the EDM matrix element can be approximated as the error in the quantity $\sqrt{A_{7S_\frac{1}{2}}A_{7P_\frac{1}{2}}}$. This can be seen from the following reasoning. The hyperfine constant $A$ is expressed like
\begin{equation}
A = \frac{\langle \Psi \vert \hat{H}_\mathrm{hf} \vert \Psi \rangle}{IJ}
\end{equation}
where $\hat{H}_\mathrm{hf}$ is the hyperfine interaction Hamiltonian:
\begin{equation}
\langle \hat{H}_\mathrm{hf} \rangle = \langle  \sum_e \bm{j}_e \cdot \bm{A}_N \rangle = A \langle \bm{I} \cdot \bm{J}\rangle .
\end{equation}
Thus,
\begin{eqnarray}
\sqrt{A_{7S_\frac{1}{2}}A_{7P_\frac{1}{2}}} &\propto& \sqrt{\langle {7S_\frac{1}{2}} \vert \hat{H}_\mathrm{hf} \vert {7S_\frac{1}{2}} \rangle \langle {7P_\frac{1}{2}} \vert \hat{H}_\mathrm{hf} \vert {7P_\frac{1}{2}} \rangle}. \nonumber 
\end{eqnarray}
The dominant contribution to the EDM matrix element is due to the transition between the $7S_\frac{1}{2}$ and $7P_\frac{1}{2}$ states, given as
\begin{equation}
\langle 7S_\frac{1}{2} \vert \beta\gamma_5{p}^2 \vert 7P_\frac{1}{2} \rangle.
\end{equation}
If we note that both $\hat{H}_\mathrm{hf}$ and $\beta\gamma_5{p}^2$ are one-body operators, and that both are sensitive to contributions from orbitals with a dominant component in the nuclear region, we can see that the accuracy of $\sqrt{A_{7S_\frac{1}{2}}A_{7P_\frac{1}{2}}}$ can give some indication of the accuracy of $\langle 7S_\frac{1}{2} \vert \beta\gamma_5{p}^2 \vert 7P_\frac{1}{2} \rangle$, and therefore of $\langle \Psi_\alpha^\mathrm{(0)}\vert \beta\gamma_5{p}^2 \vert \Psi_\nu^\mathrm{(0)}\rangle$. 
Other terms in Eq.~(\ref{eq:R-expr}) can be compared directly against experimental results, so this term is the greatest source of uncertainty in the error estimate.

Table~\ref{tb:A-all} shows the RCC values of the hyperfine constants of the 7$S_\frac{1}{2}$, 7$P_\frac{1}{2}$, 8$P_\frac{1}{2}$, and 9$P_\frac{1}{2}$ states, with various correction terms listed and the result compared against the available experimental~\cite{Grossman} and theoretical results~\cite{Sahoo2}. The included correction terms are the Breit interaction terms, the approximate QED effect terms, the perturbative triple excitation (pT) terms, and the Bohr-Weisskopf (BW) effect terms. The Breit interaction is the lowest order relativistic correction to the Coulomb interaction~\cite{Grant, Breit}, and the QED effect terms include corrections due to vacuum polarization effects and electron self-energy effects, calculated approximately~\cite{Yu, Ginges}. The correction due to inclusion of partial effective three particle-three hole excited states is evaluated by treating the excitation as a perturbation on the evaluated RCCSD state, as outlined by Sahoo et al. in Ref.~\cite{Sahoo}. The BW effect is the correction due to the magnetization of the nucleus~\cite{Bohr, Ginges2}, to which the hyperfine constant values are sensitive. For the 7$S_\frac{1}{2}$ state, our final calculation gives a value with an approximate 0.76\% deviation from experimental results, which is an excellent agreement. The value for the 7$P_\frac{1}{2}$ state also shows good agreement with experimental results, with a deviation of about 0.64\%. Our results give a marginally better agreement with available experimental results compared to the results in Ref.~\cite{Sahoo2}. The main differences between our method and the RCC method used in Ref.~\cite{Sahoo2} is that Ref.~\cite{Sahoo2} have used a quadratic basis set instead of the GTO's that we have used, and that Ref.~\cite{Sahoo2} have not included the BW contributions. Our results demonstrate the power of the RCC method used in this work to provide reliable calculations of atomic properties.
\begin{table}[h!]
\caption{RCC calculations of selected hyperfine structure constant quantities ($A$) for $^{210}$Fr using the DC Hamiltonian and the correction terms due to the Breit interaction, approximate QED effects, perturbative triple excitation (pT) terms, and the BW effect. The final results are compared against the available experimental measurements and theoretical values. Values are given in units of MHz.}

 \begin{tabularx}{8.6cm}{X X  X  X  X  X} 
 \hline
 \hline
 \multicolumn{2}{l}{Term} & 7$S_\frac{1}{2}$ & 7$P_\frac{1}{2}$ & 8$P_\frac{1}{2}$ & 9$P_\frac{1}{2}$\\ [0.5ex] 
 \hline
 \multicolumn{2}{l}{DC} & 7488.42 & 944.56 & 296.22 & 132.80 \\ 
 \multicolumn{2}{l}{Breit} & 16.217 & -1.584 & -0.363 & -0.108 \\
 \multicolumn{2}{l}{QED} & -41.026 & 3.466 & 0.970 & 0.421 \\ 
 \multicolumn{2}{l}{pT} & -14.389 & 1.959 & 0.507 & 0.205 \\
 \multicolumn{2}{l}{BW} & -199.228 & -8.163 & -2.563 & -1.149\\
 \hline
 \multicolumn{2}{l}{Total} & 7249.99 & 940.236 & 294.772 & 132.173\\
 \hline
 \hline
 \multicolumn{2}{l}{Experiment~\cite{Grossman}} & 7195.1(4) & 946.3(2) & - & - \\
 \multicolumn{2}{l}{Ref.~\cite{Sahoo2} (theory)} & 7254(75) & 939(7) & 295(4) & -\\
 \hline
 \hline
\end{tabularx}
\label{tb:A-all}
\end{table}

Table~\ref{tb:DC-A} shows the values of selected hyperfine structure constant quantities of the form $\sqrt{A_{7S_\frac{1}{2}}A_{nP_\frac{1}{2}}}$ (for $n=7,8,9$) calculated using the RCC wave function, compared against the available experimental results. Here, the results for $A$ using the DC Hamiltonian without the various correction terms are used, so that we obtain a conservative estimate. For $\sqrt{A_{7S_\frac{1}{2}}A_{7P_\frac{1}{2}}}$, the calculated value is about 1.9\% larger than the experimental value. From the argument above, this can be thought of as the largest error coming from the EDM matrix element.
\begin{table}[h!]
\caption{RCC calculations of selected hyperfine structure constant quantities for $^{210}$Fr using the DC Hamiltonian, compared against the available experimental measurements.}

 \begin{tabularx}{8.6cm}{X  X  X} 
 \hline
 \hline
 Terms & This work & Experiment\\ [0.5ex] 
 \hline
 $\sqrt{A_{7S_\frac{1}{2}}A_{7P_\frac{1}{2}}}$ & 2657.79 & 2609.37~\cite{Grossman} \\ 
 $\sqrt{A_{7S_\frac{1}{2}}A_{8P_\frac{1}{2}}}$ & 1488.43 & - \\
 $\sqrt{A_{7S_\frac{1}{2}}A_{9P_\frac{1}{2}}}$ & 996.76 & - \\ 
 \hline
 \hline
\end{tabularx}

\label{tb:DC-A}
\end{table}

The second quantity to compare is the E1 transition amplitude, found in the numerator of Eq.~(\ref{eq:R-expr}). Table~\ref{tb:DC-E1} shows the values of the E1 transition amplitudes of selected valence transitions calculated using the RCC wave function used in the enhancement factor calculation, compared against the available experimental values~\cite{Simsarian} and theoretical calculations~\cite{Safronova, Dzuba}. It can be seen that the various correction terms added to the DC result (Breit and QED) have a limited effect on the final values of the transition amplitudes. For the 7S$_\frac{1}{2} \rightarrow$ 7P$_\frac{1}{2}$ transition, the magnitude of the calculated quantity is about 1.59\% larger than the experimental value, which is again in good agreement, especially considering that the many-body wave function was optimized not just for the transition amplitudes but also simultaneously for the hyperfine constants and the EDM enhancement factor. Comparison against two other theoretical results are made in Table~\ref{tb:DC-E1} as well. Ref.~\cite{Safronova} uses a linearized coupled-cluster method, which only considers terms that are linear in the cluster operators $T$ and $S$, and Ref.~\cite{Dzuba} uses a many-body perturbation theory with screened Coulomb interactions. Ref.~\cite{Safronova} also uses linear combinations of B-splines as the basis orbitals, instead of the GTO's that we use here. Our calculation includes many more terms than what have been included in both Ref.~\cite{Safronova} and Ref.~\cite{Dzuba}.
\begin{table}[h!]
\caption{RCC calculations of magnitudes of selected E1 transition amplitudes for $^{210}$Fr using the DC Hamiltonian and the correction terms due to the Breit interaction and approximate QED effects. The final results are compared against the available experimental measurements and theoretical values. Values are given in units of Bohr radius.}
 \begin{tabularx}{8.6cm}{X  X  X  X  X} 
 \hline
 \hline
 \multicolumn{2}{l}{Term} & 7S$_\frac{1}{2} \rightarrow$ 7P$_\frac{1}{2}$ & 7S$_\frac{1}{2} \rightarrow$ 8P$_\frac{1}{2}$ & 7S$_\frac{1}{2} \rightarrow$ 9P$_\frac{1}{2}$ \\ [0.5ex] 
 \hline
 \multicolumn{2}{l}{DC} & 4.345 & 0.333 & 0.111 \\ 
 \multicolumn{2}{l}{Breit} & 0.0004 & 0.0028 & 0.0016 \\
 \multicolumn{2}{l}{QED} & -0.0005 & -0.0019 & -0.0011 \\ 
 \hline
 \multicolumn{2}{l}{Total} & 4.345 & 0.334 & 0.114 \\
 \hline
 \hline
 \multicolumn{2}{l}{Experiment~\cite{Simsarian}} & 4.277(8) & - & - \\
 \multicolumn{2}{l}{Ref.~\cite{Safronova} (theory)} & 4.256 & 0.327 & 0.110 \\
 \multicolumn{2}{l}{Ref.~\cite{Dzuba} (theory)} & 4.304 & 0.301 & - \\
 \hline
 \hline
\end{tabularx}
\label{tb:DC-E1}
\end{table}

The third is the excitation energy between the ground and selected excited states, as seen in the denominator of Eq.~(\ref{eq:R-expr}). For the selected excitation energies, all three calculation results show a very small deviation from experimental results, at 0.47\%, 0.17\%, and 0.11\%, for the 7P$_\frac{1}{2} \rightarrow$ 7S$_\frac{1}{2}$, 8P$_\frac{1}{2} \rightarrow$ 7S$_\frac{1}{2}$, and 9P$_\frac{1}{2} \rightarrow$ 7S$_\frac{1}{2}$ transitions, respectively. This is not surprising, as the DF method optimizes the orbital wave functions by minimizing the DF energy value. Thus, the errors introduced from this term can be assumed to be smaller than the errors due to the other two terms.
\begin{table}[h!]
\caption{RCC calculations of selected excitation energies for $^{210}$Fr using the DC Hamiltonian, compared against the experimental measurements given in Ref.~\cite{Sansonetti}. The values are given in units of cm$^{-1}$.} 

 \begin{tabularx}{8.6cm}{X  X  X} 
 \hline
 \hline
 Transition & This work & Experiment~\cite{Sansonetti}\\ [0.5ex] 
 \hline
 7P$_\frac{1}{2} \rightarrow$ 7S$_\frac{1}{2}$ & 12295.04 & 12237.41 \\
 8P$_\frac{1}{2} \rightarrow$ 7S$_\frac{1}{2}$  & 23151.49 & 23112.96 \\ 
 9P$_\frac{1}{2} \rightarrow$ 7S$_\frac{1}{2}$  & 27149.03 & 27118.21 \\
 \hline
 \hline
\end{tabularx}

\label{tb:DC-energy}
\end{table}
Taking these three sources of errors to be independent, we add the fractional uncertainties in quadrature, and obtain a total conservative estimated error of about 3\%. Notably, previous results do not fall within the error bar of this result, indicating that the improvements we have made in this calculation, especially the additional terms in $\bar{D}$ we have included, have a significant effect on the final result, and may indicate that continued efforts for an improved calculation are necessary. 
We emphasize here that evaluations of the errors in the individual terms in Eq.~(\ref{eq:R-expr}), by this method or otherwise, had not been performed by Mukherjee et al.~\cite{Fr1}. Byrnes et al.~\cite{Fr2} have reported estimates of the errors in excitation energies, E1 transition amplitudes, and the EDM matrix elements, and have reported a total error of 5\%. Our calculations provide a comprehensive analysis of the error in our calculation of $R$.

We now discuss the corrections introduced to $R$ due to the consideration of the Breit interaction and approximate QED effects. Table~\ref{tb:Breit-210} shows the RCC contributions to $R$ for $^{210}$Fr with Breit interaction effects accounted for. 
\begin{table}[h!]
\caption{DF and the leading RCC contributions to $R$ of the atomic EDM of $^{210}$Fr using the DC Hamiltonian with Breit interaction correction terms, calculated using the RCCSD method. ``Norm.'' refers to the correction due to normalization of the RCC wave function, and ``Extra'' refers to contributions due to terms not listed in the table, which have been calculated.}
 \begin{tabularx}{8.6cm}{X  X  X} 
 \hline
 \hline
 Terms & DC results & DC + Breit results \\ [0.5ex] 
 \hline
 DF (core) & 24.85 & 24.76 \\ 
 DF (valence) & 702.39 & 694.87 \\
 $\bar{D}T_1^{(1)} +$ h.c. & 44.05 & 43.99 \\ 
 $\bar{D}S_1^{(1)} +$ h.c. & 889.18 & 880.11 \\ 
 $\bar{D}S_2^{(1)} +$ h.c. & -49.46 & -49.07 \\
 ${S_1^{(0)}}^\dagger \bar{D} S_1^{(1)} +$ h.c. & -14.15 & -14.05 \\
 ${S_2^{(0)}}^\dagger \bar{D} S_1^{(1)} +$ h.c. & -48.45 & -48.02 \\
 ${S_1^{(0)}}^\dagger \bar{D} S_2^{(1)} +$ h.c. & -3.84 & -3.81 \\
 ${S_2^{(0)}}^\dagger \bar{D} S_2^{(1)} +$ h.c. & 11.99 & 11.90 \\
 Extra & 4.98 & 4.96 \\
 Norm. & -22.11 & -21.92 \\
 \hline
 Total & 812.19 & 804.08 \\
 \hline
 \hline
\end{tabularx}
\label{tb:Breit-210}
\end{table}
The inclusion of Breit interaction terms reduce the value of $R$ by about 8.11, or a decrease of about 1\%. This is mainly due to the difference in the $\bar{D}S_1^{(1)} +$ h.c. term. This is not surprising, as the valence 7s electron for Fr is expected to behave relativistically due to the large atomic size, and so it is natural that the Breit interaction, which is a relativistic effect, impacts terms involving the valence electron more significantly. 
Table~\ref{tb:QED-210} shows the RCC contributions to $R$ for $^{210}$Fr with approximate QED effects accounted for. 
\begin{table}[h!]
\caption{DF and the leading RCC contributions to $R$ of the atomic EDM of $^{210}$Fr using the DC Hamiltonian with approximate QED corrections, calculated using the RCCSD method. ``Norm.'' refers to the correction due to normalization of the RCC wave function, and ``Extra'' refers to contributions due to terms not listed in the table, which have been calculated.}
 \begin{tabularx}{8.6cm}{X  X  X} 
 \hline
 \hline
 Terms & DC results & DC + QED results \\ [0.5ex] 
 \hline
 DF (core) & 24.85 & 24.74 \\ 
 DF (valence) & 702.39 & 701.96 \\
 $\bar{D}T_1^{(1)} +$ h.c. & 44.05 & 43.84 \\ 
 $\bar{D}S_1^{(1)} +$ h.c. & 889.18 & 888.70 \\ 
 $\bar{D}S_2^{(1)} +$ h.c. & -49.46 & -49.42 \\
 ${S_1^{(0)}}^\dagger \bar{D} S_1^{(1)} +$ h.c. & -14.15 & -14.12 \\
 ${S_2^{(0)}}^\dagger \bar{D} S_1^{(1)} +$ h.c. & -48.45 & -48.44 \\
 ${S_1^{(0)}}^\dagger \bar{D} S_2^{(1)} +$ h.c. & -3.84 & -3.83 \\
 ${S_2^{(0)}}^\dagger \bar{D} S_2^{(1)} +$ h.c. & 11.99 & 11.99 \\
 Extra & 4.98 & 4.95 \\
 Norm. & -22.11 & -22.09 \\
 \hline
 Total & 812.19 & 811.57 \\
 \hline
 \hline
\end{tabularx}
\label{tb:QED-210}
\end{table}
It can be seen that the QED corrections reduce the value of $R$ slightly, mainly due to the reduction in the $\bar{D}T_1^{(1)} +$ h.c. and $\bar{D}S_1^{(1)} +$ h.c. terms, but overall the difference is very small, at about -0.62, or -0.076\%.

Finally, effective three particle-three hole excitation contributions were calculated using the DC Hamiltonian through a perturbative method~\cite{Sahoo}. In total, this resulted in a correction of about -4.64 from the DC result, or a -0.58\% correction, as shown in Table~\ref{tb:total-210}.

\begin{table}[h!]
\caption{$R$ of the atomic EDM of $^{210}$Fr calculated using the DF and the RCCSD methods, with various correction terms included. ``DC'' refers to the result obtained using the Dirac-Coulomb Hamiltonian, ``QED'' to approximate QED correction terms, ``Breit'' to correction terms due to the Breit interaction, and ``pT'' to the effective three particle-three hole excitation contribution terms.}
 \begin{tabularx}{8.6cm}{X X X X} 
 \hline
 \hline
 \multicolumn{2}{l}{Method} & Correction & $R$ \\ [0.5ex] 
 \hline
 \multicolumn{2}{l}{DF (DC)} & - & 727.24 \\ 
 \multicolumn{2}{l}{RCCSD (DC)} & 0 & 812.19 \\
 \multicolumn{2}{l}{RCCSD (DC+Breit)} & -8.105 & 804.08 \\ 
 \multicolumn{2}{l}{RCCSD (DC+QED)} & -0.621 & 811.57 \\ 
 \multicolumn{2}{l}{RCCSD (DC+pT)} & -4.644 & 807.55 \\
 \multicolumn{2}{l}{RCCSD (DC+Breit+QED+pT)} & -13.369 & 798.82 \\
 \hline
 \hline
\end{tabularx}
\label{tb:total-210}
\end{table}

If we combine the Breit interaction correction, the approximate QED correction, and the perturbative triples correction, we obtain a final value of $R = 799$ for $^{210}$Fr. We see from Table~\ref{tb:total-210} that the three correction terms each reduce the value of $R$ from the RCCSD DC value, leading to a smaller final value than the pure DC result. We note that our work is the first to apply all of these correction terms to the calculation of $R$ for $^{210}$Fr. 

\section{Conclusion}
Results for improved RCC calculations of the EDM enhancement factor for $^{210}$Fr are presented in this work, evaluated to be at $R = 799$, with an estimated error of about 3\%. This is about 11\% smaller than the result from a previous calculation using an approximate RCCSD method~\cite{Fr1}. This difference can be attributed to the fact that the various approximations and shortcomings in the previous calculation were addressed in this work, such as by the improvement of both the size and quality of the basis functions used, and by the inclusion of amplitudes of all multipoles and nonlinear RCC terms using a self-consistent approach, applied to an open-shell system for the first time here. 
We emphasize that we have outlined the method of error evaluation more comprehensively than what is given in previous calculations of the same atom, if given at all.
A detailed analysis of the many-body effects contributing to the EDM enhancement of Fr was given as well, and it was found that BPC and ECP effects contribute most heavily to $R$. This has shed light on the many-body physics involved in this complex phenomenon.
This work has also included corrections due to the Breit interaction and QED effects, as well as contributions due to perturbative triple excitations, which previous Fr EDM calculations have not included. 
The improved RCC method has also been used to evaluate the magnetic dipole hyperfine constants and the E1 transition amplitudes of selected states of $^{210}$Fr and compared against available experimental and theoretical results, with relevant correction terms included. Our results showed excellent agreement with both experiment and other detailed theoretical calculations, and have demonstrated the versatility of our RCC method in providing accurate results for atomic properties. 

The notable difference between our results with previous results indicates that it is necessary to continue to perform these calculations to higher levels of accuracy, for a reliable appraisal of the upper limit of the magnitude of the electron EDM in conjunction with future experimental results. For a comprehensive study of BSM physics, it is necessary to conduct EDM searches on multiple candidates. In this respect, the continuation of efforts for EDM measurement in atomic systems is still important. The merit of Fr in particular as an electron EDM search candidate lies in its large enhancement factor and its ability for the theoretical calculation of its $R$ to be obtained to a higher accuracy compared to molecules. In addition, it is an ideal system in which to also probe the S-PS  coupling constants, which also contribute to the atomic EDM, because of the wealth of isotopes available for production. We hope that our theoretical work on this promising candidate will not only complement experimental results, but also contribute to the understanding of relativistic many-body theory in atoms, and the development of RCC methods in general.

\section{Acknowledgements}
The calculations in this work were performed on the TSUBAME 3.0 supercomputer at the Tokyo Institute of Technology, through the TSUBAME Grand Challenge Program. 
The authors would like to thank Professor Y. Sakemi and Dr. K. Harada for providing useful information on experimental aspects of Fr EDM, and Dr. V. A. Dzuba for providing useful information related to theoretical calculations of Fr EDM.

\end{document}